\documentclass[12pt]{article}

\newcommand \qed {\vrule height5pt width5pt}

\newcommand \no {\noindent}
\newcommand \ra {\rightarrow}

\newcommand \s {\sigma}
\newcommand \half {{1 \over 2}}

\newcommand \eps {\epsilon}
\newcommand \e {\epsilon}

\newcommand \h {{\exp^{(2)}}}
\newcommand \f { {1 \over n !}}
\newcommand \weight {(|\epsilon| M)}
\def \d {\triangle}
\def \L {\Lambda}
\def \xx {{\cal X}}

\def \zed {Z}

\newcommand \fp {F} 

\newcommand{\ba}[1]{\begin{array}{#1}}
\newcommand{\ea}{\end{array}}

\newcommand{\be}{\begin{equation}}
\newcommand{\ee}{\end{equation}}
\newcommand{\bea}{\begin{eqnarray}}
\newcommand{\eea}{\end{eqnarray}}
\newcommand{\beann}{\begin{eqnarray*}}
\newcommand{\eeann}{\end{eqnarray*}}

\newcommand{\nn}{\hbox{nn}}

\def\reff#1{(\ref{#1})}

\newtheorem{theorem}{Theorem}

\begin{document}

\title{A fixed point equation for one quasiparticle states in quantum 
spin--1/2 systems}

\title{Expansions for one quasiparticle states in spin 1/2 systems}

\author{Nilanjana Datta 
\\Statistical Laboratory
\\Centre for Mathematical Sciences
\\University of Cambridge
\\Wilberforce Road, Cambridge CB30WB
\\ email: n.datta@statslab.cam.ac.uk
\\
\\Tom Kennedy
\\Department of Mathematics
\\University of Arizona
\\Tucson, AZ 85721
\\ email: tgk@math.arizona.edu
\bigskip
}

\maketitle

\begin{abstract}
Convergent expansions of the wavefunctions for the ground state
and low-lying excited states of quantum transverse Ising systems
are obtained. These expansions are employed to prove that 
the dispersion relation for a single quasi-particle 
has a convergent expansion.
\end{abstract}

\newpage

\section{Introduction} 

The ground state and low-lying 
excited states of a quantum spin system, governed by
a Hamiltonian $H$,  are usually studied by considering a low temperature
expansion of its partition function e.g. \cite{alb, ginibre, ty1, ty2}. 
This involves an analysis 
of the trace of the 
operator, $\exp(-\beta H)$, in the limit of large $\beta$ 
(with $\beta = 1/k_B T$, $k_B$ being the Boltzmannn constant). 
The analysis is usually done by applying the method of cluster 
expansions to the Trotter product representation of the operator. 
In contrast, Kirkwood and Thomas \cite{kt} studied the ground state 
of such systems by working directly with the Schrodinger equation, 
$H |\Psi\rangle = E |\Psi\rangle$. 
They considered a class of Hamiltonians of quantum spin-$1/2$ 
systems, which 
were small perturbations of classical Hamiltonians.
By making a clever ansatz for the form of the ground state, they 
were able to develop a convergent expansion for the ground state which 
they used to show that the infinite volume limit of the ground state
exists and has exponential decay of the truncated correlation functions. 

The method of Kirkwood and Thomas has been employed, in the $C^*$-algebraic
framework, to establish the uniqueness of translationally invariant ground 
states in quantum spin-$1/2$ systems \cite{matsuib} 
and has also been extended to study 
the ground states of models of arbitrary (though finite) spin, 
e.g. the quantum Potts model \cite{matsuia}.  

The approach of Kirkwood and Thomas can, however, be applied 
only to a restricted
class of Hamiltonians, namely to ones that satisfy a Perron-Frobenius
condition. 
In this paper, we greatly extend the class of Hamiltonians to which 
the Kirkwood-Thomas approach can be applied by removing the requirement 
of this condition.
We also simplify their method -- the simplification 
being in the estimates which show that the expansions 
converge. 
The methods of this paper are applicable to any lattice Hamiltonian of the 
form
\be
H= - \sum_{i} \s^x_i + \e \sum_{X} V_X,
\label{eq_gen_ham}
\ee
where $V_X$ is a local operator that acts only on the spins at sites in 
$X$. There is a finite constant $R$ such that $X$ is only summed over 
subsets with $g(X) <R$, where $g(X)$ is the number of bonds in the 
smallest connected set of sites containing $X$. 
Kirkwood and Thomas required that the matrix representing the local
operator, $V_X$, satisfies a Perron-Frobenius condition; 
we do not need such a condition. 

Throughout this paper we will work with a specific 
example of such a Hamiltonian, the Ising model in a transverse field.  
\be
H = - \sum_i \s^x_i + \e \sum_{<ij>} \, \s^z_i \s^z_j,
\label{eq_transverse_ising}
\ee
(The sum over $<ij>$ is over all nearest neighbor sites in the lattice.)
Usually perturbation theory in $\epsilon$ is done with the unitarily
equivalent Hamiltonian
\be
\widetilde{H} = - \sum_i \s^z_i + \e \sum_{<ij>} \, \s^x_i \s^x_j,
\label{eq_orig_ham}
\ee
However, following Kirkwood and Thomas, we use 
\reff{eq_transverse_ising} in our approach. (When $\epsilon=0$ the 
groundstate of  \reff{eq_transverse_ising}, in the usual basis in 
which the $\s^z$ are diagonal, is the constant wave function. 
The Kirkwood-Thomas approach is an expansion around this constant
wavefunction.) 
An example of a Hamiltonian to which our methods apply, but the original 
approach of Kirkwood and Thomas does not, is the following.
\be
H = - \sum_{i} \s^x_i + \e \sum_{<ij>} \, [\s^z_i \s^z_j + J \s^y_i \s^y_j],
\ee
with $J \ne 1$.  

A Hamiltonian of the form 
\be
H= - \sum_{i} \s^z_i + 
\e \sum_{X } V_X^\prime
\ee
is unitarily equivalent to a Hamiltonian of the form \reff{eq_gen_ham}.
A more general class of Hamiltonians one might consider is 
\be
H= \sum_X D_X + \e \sum_Y \, V_Y^\prime,
\ee
where $D_X$ is a local {\it diagonal} operator.
If the classical Hamiltonian $\sum_X D_X$ has a unique ground state,
then $Tr(e^{-\beta H})$, and hence the ground state,  
may be studied by standard expansion methods. It would be interesting 
to extend the methods of Kirkwood and Thomas and this paper to this
class of models. 

It is of interest to study, not only the properties of the ground state,
but also the low-lying excited states of quantum spin systems.
There have been a series of papers \cite{zkm,amz, ms, km, pokorny, mm} 
devoted to the investigation
of the low-lying spectra of infinite-dimensional quantum lattice systems.
In \cite{zkm} this has been done by reducing the problem to the spectral 
analysis of an operator which is unitarily equivalent to the Hamiltonian
governing the lattice system, and is also 
the generator of an infinite-dimensional
Markov diffusion process. The method of cluster expansions has been employed
to analyse the low--lying spectral properties. In particular, the spectrum of
the Hamiltonian has been analysed in the one-particle invariant subspaces, 
which describe the elementary excitations of the ground state. In \cite{amz,ms}
it has been shown that the one--particle excitations form a segment of 
absolutely continuous spectrum, which is separated by energy gaps from the 
ground state eigenvalue, and from the rest of the spectrum.

In the second part of this paper, we study one quasi-particle excitations 
in a quantum spin-$1/2$ 
system by employing a method different from the ones mentioned
above. Our method uses techniques, based on the Kirkwood-Thomas 
approach, that we develop for the ground state analysis. 
The key idea is that the lowest excited states have a definite momentum. 
We make an ansatz for the form
of these states based on their momentum, and use it to develop a 
convergent expansion for them. As an application of this expansion 
we study the dispersion relation for a single quasi-particle and 
show that it has a convergent expansion.

Kirkwood and Thomas also considered models with two ground 
states related by the global spin flip, e.g., 
the highly anisotropic $XY$- and Heisenberg models
\cite{kt}. The simplifications and generalizations of their method
that we introduce in this paper also apply to these types of models
as will be explained elsewhere \cite{dk}. 
We will also show that 
interfaces (kink states) in these models may by studied in much the same way 
that we study the excited states with one quasiparticle in the models
of this paper. 

We consider the Hamiltonian 
to be restricted to finite volumes, $\Lambda$, which are hypercubes, 
and impose periodic boundary conditions. 
This restriction is essential for our treatment of the one quasi-particle
excited states which relies on the translation symmetry of the model.
However, for our treatment of the ground state we could consider 
more general $\Lambda$ and other boundary conditions.

It has been shown \cite{pokorny} that the spectrum of the Hamiltonian
\reff{eq_transverse_ising} above the lowest energy level 
contains a {\em{continuous}} 
part and is separated from the lowest level by an energy gap. In this
paper we make no further analysis of the {\em{nature}} of the spectrum
above the ground state eigenvalue.

Section \ref{ground1} of this paper is devoted to our version of
the Kirkwood-Thomas approach to the ground state of quantum
spin-1/2 systems, in particular to the Hamiltonian in 
\reff{eq_transverse_ising}. 
We prove that the wavefunction of the Hamiltonian can be expressed in the 
form 
\be 
\psi(\s)=\exp(\sum_X h_X(\s)) 
\ee 
where $\sum_X h_X$ is a function of the configuration on the lattice
which can be thought of as a {\it classical} Hamiltonian.
It is infinite range ($h_X \ne 0$ for arbitrarily large $X$), but
the size of $h_X$ decays exponentially as $X$ grows.
This is the form of the Kirkwood-Thomas ansatz for the wavefunction, but
we emphasize that our approach shows that it holds even for models 
without a Perron-Frobenius condition. (Depending on the model, the 
function $\sum_X h_X$ may be complex-valued.)
The one-particle excitations of the system are 
analysed in Section \ref{excited1}. 
We obtain an expression for the wavefunction of the
lowest excited state in a subspace of fixed momentum,
and use it to prove that the infinite volume
dispersion relation for the spectral gap above the ground state
eigenvalue has a convergent expansion. This shows that 
this gap persists in the infinite-volume limit and allows us to 
prove properties of the gap, e.g., that it is minimized at a particular
value of $k$, depending on the sign of $\epsilon$. 
Unfortunately, the analysis of two-particle excitations lies
outside the scope of our method.

\section{The ground state}
\label{ground1}

A classical spin configuration on the lattice is an assignment
of a $+1$ or a $-1$ to each site in the lattice. So for each
$i \in \L$, $\s_i = \pm 1$. We will abbreviate the classical
spin configuration $\{\s_i\}_{i \in \L}$ by $\s$ .
For each such $\s$ we let $|\s>$ be the 
state in the Hilbert space which is the tensor product 
of a spin up state at each site with $\s_i=+1$ and a spin down state
at each site with $\s_i=-1$. 
Thus $|\s>$ is an  eigenstate of all the $\s^z_i$ with 
$\s^z_i |\s> = \s_i |\s>$.
Any state $|\Psi>$ can be written in terms of this basis:
\be
|\Psi> = \sum_\s \psi(\s) |\s>,
\label{eqa}
\ee
where $\psi(\s)$ is a complex-valued function on the spin configurations
$\s$.

The form of the Hamiltonian in equation \reff{eq_orig_ham}
seems natural for perturbation
theory in $\e$ since the Hamiltonian is diagonal when $\e=0$. 
However, following Kirkwood and Thomas, we 
study a unitarily equivalent Hamiltonian
\be
H= - \sum_i \s^x_i + \e \sum_{<ij>} \, \s^z_i \s^z_j.
\label{eqb}
\ee
For a single site the ground state of $-\s^x$ is the vector
$|+1>+|-1>$. Thus the unnormalized
ground state of the Hamiltonian \reff{eqb} with $\e=0$ is
given by \reff{eqa} with $\psi(\s)=1$ for all $\s$. 

For a subset $X$ of the lattice $\L$, let 
$$\s(X) = \prod_{i \in X} \s_i,$$
and we use the convention that $\s(\emptyset) =1.$
Let $f(\s)$ be a complex-valued function of the $\s$'s. 
Then $f(\s)$ can be written in a unique way as 
$$f(\s)= \sum_{X} g(X) \s(X),$$
where the $g(X)$ are complex numbers and $X$ is summed over 
all subsets of $\L$ (including $\L$ itself and the empty set).
The Fourier coefficients $g(X)$ are given by
$$g(X) = 2^{-|\L|} \, \sum_\s \, \s(X) \, f(\s),$$
where $|\L|$ is the number of sites in $\L$ and 
the sum is over all $2^{|\L|}$ spin configurations on $\L$. 
[In general, let $|X|$ denote the number of sites in any 
subset $X$ of the lattice.]

The heart of the Kirkwood Thomas method is the following.
We expand the ground state with respect to the basis as 
in eq. \reff{eqa} and 
assume that $\psi(\s)$ can be written in the form 
\be
\psi(\s)= \exp[- {1 \over 2} \sum_X g(X) \s(X)] .
\label{eqc}
\ee
(The factor of $-{1 \over 2}$ is introduced for later convenience.)
Kirkwood and Thomas use a Perron-Frobenius argument to show that
$\psi(\s)>0$ for all $\s$, and so the ground state can be written in 
this form with real $g(X)$. 
The Perron-Frobenius argument requires that the off-diagonal terms
in the Hamiltonian are all non-positive, and so it appears that
the Kirkwood Thomas approach only applies to a very restricted 
class of models. This is not the case. We will show how to 
eliminate the Perron-Frobenius argument entirely.
We proceed in two steps. First, we will make the
ansatz \reff{eqc} and prove that there is an eigenstate of this form.
Later we will give a short argument to show that this eigenstate 
must in fact be the ground state.

Consider the eigenvalue equation 
\be
H |\Psi> =  E_0 |\Psi> .
\label{eqbb}
\ee
The operator $\s^z_i \s^z_j$ is diagonal, so 
\be
\s^z_i \s^z_j \sum_\s \psi(\s) |\s> 
= \sum_\s \s_i \s_j \psi(\s) |\s> .
\ee
The operator $\s^x_i$ just flips the spin at site $i$, i.e.,
$\s^x_i |\s> = |\s^{(i)}>$, where $\s^{(i)}$ is the spin configuration
$\s$ but with $\s_i$ replaced by $-\s_i$. 
Hence
\be
\s^x_i \sum_\s \psi(\s) |\s> 
= \sum_\s \psi(\s) |\s^{(i)}>
= \sum_\s \psi(\s^{(i)}) |\s> .
\ee
The last equality follows by a change of variables in the sum.

Thus if we 
use \reff{eqa} in the Schrodinger equation \reff{eqbb} and pick out the 
coefficient of $|\s>$, then for each spin configuration $\s$ we have
\be
 - \sum_i \psi(\s^{(i)}) + \e \sum_{<ij>} \, \s_i \s_j \psi(\s)
= E_0 \psi(\s).
\ee
Dividing both sides by $\psi(\s)$, we have 
\be
- \sum_i { \psi(\s^{(i)}) \over \psi(\s)}
 + \e \sum_{<ij>} \, \s_i \s_j = E_0 .
\ee
Now $\s^{(i)}(X)$ is just $\s(X)$ if $i \notin X$ and  
$-\s(X)$ if $i \in X$.
Thus
$$\psi(\s^{(i)}) =\exp[- {1 \over 2} \sum_{X: i \notin X} g(X) \s(X)
+ {1 \over 2} \sum_{X: i \in X} g(X) \s(X)],
$$
and so we have 
$${\psi(\s^{(i)}) \over \psi(\s)}
 = \exp[\sum_{X: i \in X} g(X) \s(X)].
$$
Thus the Schrodinger  equation is now
\be
- \sum_i \exp[\sum_{X: i \in X} g(X) \s(X)]
+ \e \sum_{<ij>} \, \s_i \s_j = E_0. 
\label{eqd}
\ee
This equation must hold for every configuration $\s$.
This is the magic step in the Kirkwood Thomas method because of the
cancellation between terms in $\psi(\s)$ and $\psi(\s^{(i)})$.
As $|\L| \rightarrow \infty$, 
the sum $\sum_X g(X) \s(X) $ will diverge,
but the sum $\sum_{X: i \in X} g(X) \s(X)$ that appears in the
above equation will remain bounded.
We will refer to equation \reff{eqd} as the Kirkwood Thomas equation.

We define
\be
\h (x) = e^x - 1 - x.
\ee
and rewrite \reff{eqd} as 
\bea
 - \sum_i \exp^{(2)}\bigg(\sum_{X: i \in X} g(X) \s(X)\bigg)
+ \e \sum_{<ij>} \, \s_i \s_j &=& E_0 + |\L|
+  \sum_i \sum_{X: i \in X} g(X) \s(X) \nonumber\\
&=& E_0 + |\L| + \sum_X |X| g(X) \s(X)
\eea
So the Kirkwood Thomas equation becomes
\be
\sum_X |X| g(X) \s(X) + E_0 + |\L|
= 
 - \sum_i \h \bigg(\sum_{X: i \in X} g(X) \s(X)\bigg)
+ \e \sum_{<ij>} \, \s_i \s_j.
\ee

Expanding $\h$ in a power series we have
\bea
\sum_X |X| g(X) \s(X) + E_0 + |\L|
&=& 
 - \sum_i \sum_{n=2}^\infty \, \f \bigg(\sum_{X: i \in X} g(X) \s(X)\bigg)^n
+ \e \sum_{<ij>} \, \s_i \s_j \nonumber\\
&=& 
 - \sum_i \sum_{n=2}^\infty \, \f 
\sum_{X_1,X_2,\cdots,X_n: i \in X_j}
\prod_{j=1}^n  g(X_j) \s(X_j)
+ \e \sum_{<ij>} \, \s_i \s_j.\nonumber\\
\label{kt}
\eea
The notation $: i \in X_j$ means that $i$ belongs to $X_j$ for all 
$j=1,2,\cdots,n$. 
Since $\s_i^2=1$, $\s(X) \s(Y) = \s(X \d Y)$ where the
symmetric difference $X \d Y$ of $X$ and $Y$ is defined by
$X \d Y = (X \cup Y) \setminus (X \cap Y)$. 
Thus
$\prod_{j=1}^n \s(X_j) = \s(X_1 \d \cdots \d X_n).$ 
Eq. \reff{kt} implies that the coefficients of $\s (X)$ on both sides of 
the equation must be equal. So for $X \ne \emptyset$ we have
\be
g(X) =  - { 1 \over |X|} \sum_i \sum_{n=2}^\infty \, \f 
\sum_{X_1,X_2,\cdots,X_n: i \in X_j, 
\atop{X_1 \d \cdots \d X_n=X} } 
g(X_1) g(X_2) \cdots g(X_n) 
\, + \, {1 \over |X|} \e \, 1_{\nn}(X), 
\label{fixedpt}
\ee
where $1_{\nn}(X)$ is $1$ if $X$ consists of two nearest neighbor 
sites and is $0$ otherwise.

Similarly, for $X=\emptyset$, we obtain the equation
\be
E_0 + |\L| =  - \sum_i \sum_{n=2}^\infty \, \f 
\sum_{X_1,X_2,\cdots,X_n: i \in X_j, 
\atop{X_1 \d \cdots \d X_n= \emptyset} } 
g(X_1) g(X_2) \cdots g(X_n),
\ee
since $\s(\emptyset) = 1$.

We let $g$ denote the collection of coefficients 
$\{g(X): X \subset \Lambda, X \ne \emptyset\}$,
and think of eq. \reff{fixedpt} as a fixed point equation, $g=\fp(g)$.  
We define a norm by 
\be
||g||= \sum_{X \ni i} \, |g(X)| \, |X|\, \weight^{-w(X)} 
\label{normg}
\ee
where $w(X)$ is the number of bonds in the smallest connected set
of bonds that contains $X$.
(We make the convention that a bond contains its endpoints.)
If $X$ just contains the single site $i$, then $w(X)=0$. 
With periodic boundary conditions we need only consider 
$g(X)$ which are translation invariant, and so 
the above quantity does not depend on $i$.
With other boundary conditions we would need to define the norm to 
be the supremum over $i$ of the above. 
The constant $M$ will be chosen later. It will only depend on the lattice.
We will prove that the fixed point equation has a solution in the 
Banach space defined by this norm if $|\epsilon|$ is sufficiently small
by applying the contraction mapping theorem. 

\begin{theorem}
There exists a constant $M>0$ which depends only on the number of 
dimensions such that if $|\eps| M \le 1$,
then the fixed point equation \reff{fixedpt} has a solution $g$, 
and $||g|| \le c $ for some constant $c$ which depends 
only on the lattice.
\label{thmone}
\end{theorem}

\no {\bf Proof:} We will prove that $F$ is a contraction on 
a small ball about the origin, and that it maps this 
ball back into itself. The contraction mapping theorem
will then imply that $F$ has a fixed point in this ball. 
We denote the radius of the ball by $\delta$ and let 
\be 
\delta = { 4d \over M}.
\ee 
(The factor of $4d$ is for the lattice $\zed^d$. In general this 
factor is twice the coordination number of the lattice.)
For the sake of concreteness, we will prove $F$ is a contraction 
with constant $1/2$, but there is nothing special about the choice of 
$1/2$. So we need to show 
\be
||\fp(g)-\fp(g')|| \le \frac{1}{2} ||g - g'|| \quad {\hbox{for}} \quad
||g||, ||g'|| \le \delta , 
\label{boundone}
\ee
and 
\be
||\fp(g)|| \le \delta \quad {\hbox{for}} \quad
||g|| \le \delta . 
\label{boundtwo}
\ee

The proof of \reff{boundone} proceeds as follows.
\be
||\fp(g)-\fp(g')|| \le 
\sum_i \sum_{n=2}^\infty {1 \over n!}
\sum_{X_1,\cdots,X_n: i \in X_j, 
\atop{0 \in X_1 \d \cdots \d X_n} } 
|g(X_1) \cdots g(X_n) - g'(X_1) \cdots g'(X_n)| 
\weight^{-w(X_1 \d \ldots \d X_n)} .
\label{return}
\ee
If $0 \in X_1 \d \ldots \d X_n$, then $0$ is in at least one $X_j$. 
Using the symmetry under permutations of the $X_j$, we can 
take $0 \in X_1$ at the cost of a factor of $n$. 
We claim that if $i \in X_j$ for $j=1,2,\cdots,n$, then 
\be
w(X_1 \d \cdots \d X_n) \le \sum_{j=1}^n w(X_j) .
\label{wtri}
\ee
To prove the claim, 
let $C_j$ be connected sets of bonds such that 
$X_j \subset C_j$ and $|C_j|=w(X_j)$.
Define $C=\cup_{j=1}^n C_j$. Since all the $X_j$ contain $i$, all the 
$C_j$ contain $i$. So $C$ is connected. Clearly, 
$X_1 \d \cdots \d X_n \subset C$.
So 
\be w(X_1 \d \cdots \d X_n) \le |C| \le \sum_{j=1}^n |C_j|
= \sum_{j=1}^n w(X_j)
\ee
which proves the claim \reff{wtri}.
So  
\bea
&& ||\fp(g)-\fp(g')|| 
\nonumber\\
&& \le
\sum_i \sum_{n=2}^\infty {1 \over (n-1)!}
\sum_{X_1, \cdots, X_n: i \in X_j, 0 \in X_1} 
\left|\prod_{j=1}^n g(X_j) - 
\prod_{j=1}^n g'(X_j) \right| \weight^{-w(X_1)-w(X_2) \cdots - w(X_n)}
\nonumber\\
&& \le
\sum_{n=2}^\infty {1 \over (n-1)!}
\, \sum_{X_1 \ni 0} \, \sum_{i \in X_1} \, \sum_{X_2 \ldots X_n \ni i} 
\, \sum_{k=1}^n \bigg( \prod_{j=1}^{k-1} |g(X_j)| \bigg) |g(X_k)-g'(X_k)| 
\bigg( \prod_{j=k+1}^n |g'(X_j)| \bigg) \nonumber\\
&& \quad \quad \quad \times \quad 
\weight^{-w(X_1) - w(X_2) \cdots - w(X_n)} \nonumber\\
&& \le  \sum_{n=2}^\infty {1 \over (n-1)!}
\, \sum_{X_1 \ni 0} \, \sum_{i \in X_1} |g(X_1)-g'(X_1)| \weight^{-w(X_1)}
\sum_{X_2 \ldots X_n \ni i} \, \bigg( \prod_{j=2}^{n} |g'(X_j)|
\weight^{-w(X_j)} \bigg)
\nonumber\\
&& + \sum_{n=2}^\infty {1 \over (n-1)!}
\, \sum_{X_1 \ni 0} 
\, \sum_{i \in X_1} |g(X_1)|\weight^{-w(X_1)}
\, \sum_{X_2 \ldots X_n \ni i} \, \sum_{k=2}^n 
\, \bigg( \prod_{j=2}^{k-1} |g(X_j)|\weight^{-w(X_j)} \bigg)
\nonumber\\
&& \quad \quad \times \quad |g(X_k)-g'(X_k)| \weight^{-w(X_k)}
\, \bigg( \prod_{j=k+1}^{n} |g'(X_j)| \weight^{-w(X_j)} \bigg)
\nonumber\\
&& \le  || g - g'|| \sum_{n=2}^\infty {1 \over (n-1)!}
\left[ ||g'||^{n-1} + \sum_{k=2}^n ||g||^{k-1} \, ||g'||^{n-k}\right],
\label{g11}
\eea
where we have bounded various sums by $||g-g'||$ ,$||g||$ and $||g'||$
in obvious ways.
Since $||g||$ and $||g'||$ are both bounded by $\delta$, 
it follows from \reff{g11} that 
\be
||\fp(g) - \fp(g')||\le K \, ||g - g'|| ,
\label{f1}
\ee
where
\be
K = \sum_{n=2}^\infty {n \over (n-1)!} \delta^{n-1}
= e^{\delta} - 1 + \delta \, e^{\delta}.
\label{kg}
\ee
Recalling that $\delta=4d/M$, 
if $M$ is large enough then $K \le 1/2$.

To prove \reff{boundtwo}, we use \reff{boundone} with $g'=0$. 
Using \reff{fixedpt} to compute $\fp(0)$, it follows that
\bea
||\fp(g)|| &\le& ||\fp(g)-\fp(0)|| + ||\fp(0)||
\le {1 \over 2} ||g|| + \sum_{X \ni 0} |\eps| 1_{\nn} (X)
  \weight^{-w(X)}\nonumber\\
& \le & {1 \over 2} ||g|| + 2 d |\eps| \weight^{-1} \le {\delta \over 2} 
+ 2 d M^{-1} = \delta .
\label{zero}
\eea
\qed

At this point all we have shown is that we can find an eigenfunction of 
the form \reff{eqc}. 
Kirkwood and Thomas show that this eigenfunction is 
the ground state by a Perron-Frobenius argument. This limits their method to 
models in which the off-diagonal matrix elements of the Hamiltonian 
are non-positive. We give a different argument which does not 
impose any restrictions on the signs of the matrix elements of the 
Hamiltonian.

When $\eps=0$, we have $g=0$, and 
we know that our eigenfunction is the ground state. 
Since we are working on a finite lattice, we are considering a finite 
dimensional eigenvalue problem. So the only way our eigenvalue can cease 
being the ground state energy is if it crosses another eigenvalue.
So if we can show that our eigenfunction is nondegenerate for all 
$\eps$ with $|\eps|< 1/M$, then it follows that our eigenfunction 
is the ground state for all such $\eps$. Suppose that there is 
some $\eps$ such that the degeneracy of the eigenvalue of our eigenfunction
is greater than one. Recall that our eigenfunction is denoted by 
$\psi(\s)$. Let $\phi(\s)$ denote a different eigenfunction with the 
same eigenvalue. Define
\be
\psi_\alpha(\s)=\psi(\s) + \alpha \phi(\s) .
\ee
So $\psi_\alpha$ is also an eigenfunction with the same eigenvalue. 
Since $\psi(\s)>0$ for all $\s$, and there are finitely many 
$\s$, if $\alpha$ is sufficiently small, then 
$\psi_\alpha(\s)>0$ for all $\s$. Thus it can be written in the form
\be
\psi_\alpha(\s)= \exp[- {1 \over 2} \sum_X g_\alpha(X) \s(X)]  .
\ee
Moreover, as $\alpha \rightarrow 0$, $||g-g_\alpha|| \rightarrow 0$. 
Since $\psi_\alpha$ is an eigenfunction, $g_\alpha$ satisfies the fixed point
equation. 
For sufficiently small $\alpha$, $g_\alpha$ is in the ball of radius 
$\delta$ about the origin. But this contradicts the uniqueness of the 
solution of the fixed point equation in this ball.
Thus the eigenvalue is nondegenerate, and so the eigenfunction we have
constructed is indeed the ground state.
In the preceding argument, how small $\alpha$ must be depends on 
$\Lambda$. This is not a problem. For each $\Lambda$ we need only 
find one value of $\alpha$ which leads to a contradiction and so implies
that the eigenfunction is not degenerate. This argument works for all 
$\epsilon$ with $|\epsilon| < 1/M$, and $M$ does not depend on $\Lambda$.

Finally, we consider the question of convergence to the infinite 
volume limit. One approach would be to show that the coefficients 
$g(X)$ have convergent expansions in $\epsilon$ and that these 
expansions agree up to order $\epsilon^L$ where $L$ is the length 
of the smallest side in $\Lambda$. 
We will sketch a different approach based on the 
fixed point equation. It takes advantage of the fact that if we have a
$g$ which is near the origin and very close to being a fixed point, 
then it must be very close to the true fixed point. 

We start by asking where the volume dependence is 
in the fixed point equation \reff{fixedpt}.
The $X_i$ in this equation must be subsets of the volume $\Lambda$, and 
the notion of nearest neighbor in the term $1_{\nn}(X)$ in  
\reff{fixedpt} depends on the volume because of the periodic boundary
conditions.
To make the volume dependence of $F(g)$ explicit, we write it as
$F_\Lambda(g)$. We can define a fixed point equation which should give
the infinite volume limit directly. 
To do this,  in \reff{fixedpt} we let the $X_i$ be subsets of 
$\zed^d$ and interpret $1_{\nn}(X)$ to mean nearest neighbors in 
$\zed^d$ in the usual sense.
We denote the resulting function by $F_\infty(g)$. Our proof of 
Theorem \ref{thmone} also shows 
that there is a small neighborhood of the origin in which the
fixed point equation $F_\infty(g)=g$ has a unique solution. 

Let $g_\Lambda$ and $g_\infty$ denote the fixed points of $F_\Lambda$ and 
$F_\infty$. We want to show that as the volume $\Lambda$ converges to 
$\zed^d$, $g_\Lambda$ converges to $g_\infty$. 
Note that $g_\Lambda$ and $g_\infty$ are defined for different collections
of subsets, so it is not immediately clear in what sense 
$g_\Lambda$ should converges to $g_\infty$. 
In fact, if a set $X$ is near the boundary, $g_\Lambda(X)$ and $g_\infty(X)$
need not be close. 
Consider one dimension and let 
$\Lambda=\{1,2,\cdots,L\}$, and let $X=\{1,L\}$. Because of the periodic
boundary conditions, the two sites in $X$ are actually nearest 
neighbors. So $g_\Lambda(X)$ is of order $\epsilon$. But $g_\infty(X)$
will be of order $\epsilon^L$. 

To compare $g_\Lambda$ and $g_\infty$ correctly, we proceed as follows.
We take our volume to be centered about the origin and let $L$  
be the length of the shortest side. So the closest sides are a distance
$L/2$ from the origin. Recall that $w(X)$ is the number of bonds 
in the smallest set of bonds which contains $X$.
We claim that 
\be
\sum_{X \ni 0: w(X) \le L/4} |g_\Lambda(X) - g_\infty(X)| \label{expfast}
\ee
converges to zero exponentially fast as $\Lambda$ converges to $\zed^d$.
Note that the terms we are neglecting, 
\be 
\sum_{X \subset \Lambda: 0 \in X, w(X) > L/4} |g_\Lambda(X)|
\ee
and 
\be 
\sum_{X \subset \zed^d: 0 \in X, w(X) > L/4} |g_\infty(X)|,
\ee
are both of order $\weight^{L/4}$ by our bounds on the solution of the 
fixed point equations.  

Our strategy is to show that $g_\Lambda$ is almost a fixed point 
of $F_\infty$. This then implies that $g_\Lambda$ is close to the exact 
fixed point $g_\infty$ of $F_\infty$. 
To show that $g_\Lambda$ is almost a fixed point 
of $F_\infty$ we need to estimate
\be
F_\infty(g_\Lambda) - g_\Lambda = F_\infty(g_\Lambda) - F_\Lambda(g_\Lambda) .
\ee
However, $g_\Lambda$ is only defined on subsets of $\Lambda$, 
so we cannot plug $g_\Lambda$ directly into $F_\infty$. 
We define a function $g^\prime_\Lambda$ which is defined for all 
$X \subset \zed^d$ as follows. For sets $X$ with $0 \in X$ and 
$w(X) \le L/4$, we let $g^\prime_\Lambda(X)=g_\Lambda(X)$.
We then extend the definition to sets which are translates of such $X$ 
by taking $g^\prime_\Lambda(X+t)=g^\prime_\Lambda(X)$.
(These translations are done in $\zed^d$.)
Finally we define $g^\prime_\Lambda(X)$ to be zero for the sets for which 
it is not yet defined. Note that there is a consistency issue here.
If we have two sets $X$ and $Y$ which both contain $0$, have 
$w(X),w(Y) \le L/4$ and are translates of each other in $\zed^d$, then 
we need to be sure $g_\Lambda(X)=g_\Lambda(Y)$. These conditions on $X$ and 
$Y$ ensure that they are also translates in $\Lambda$, so this follows 
from the translation invariance of $g_\Lambda$. 

We will now show that 
$g^\prime_\Lambda$ is close to $g_\infty$ by showing that  
$g^\prime_\Lambda$ is almost a fixed point of $F_\infty$.
We will estimate 
\be
\sum_{X \ni 0} |F_\infty(g^\prime_\Lambda)(X) - g^\prime_\Lambda(X)| 
\label{fixdif}
\ee
where $F_\infty(g^\prime_\Lambda)(X)$ is the coefficient of 
$\s(X)$ in $F_\infty(g^\prime_\Lambda)$. 
Note that we are now using a norm without the weighting factor of 
$\weight^{-w(X)}$.
We divide the terms in this sum into two classes, those with 
$w(X) \le L/4$ and those with $w(X) > L/4$.
For the latter, $g^\prime_\Lambda(X)=0$ and 
\be
\sum_{X \ni 0: w(X)>L/4} 
|F_\infty(g^\prime_\Lambda)(X)| \le  
\weight^{L/4} \sum_{X \ni 0: w(X)>L/4} 
|F_\infty(g^\prime_\Lambda)(X)| \weight^{-w(X)}
= \weight^{L/4} ||F_\infty(g^\prime_\Lambda)||
\label{latter}
\ee
where $||F_\infty(g^\prime_\Lambda)||$ is the original norm with the 
exponential weight.
Now consider an $X$ containing $0$ for which $w(X) \le L/4$. Then 
$g^\prime_\Lambda(X)=g_\Lambda(X)=F_\Lambda(g_\Lambda)(X)$.  
So we need to compare $F_\infty(g^\prime_\Lambda)(X)$ and 
$F_\Lambda(g_\Lambda)(X)$.  
Consider a term $X_1,\cdots,X_n$ in $F_\infty$ with $\sum_i w(X_i) \le L/4$
and $X_1 \d \cdots \d X_n = X$, {with $X$ containing the origin}.
If $w(X_i)\le L/4$, then $X_i$ has a translate $X_i+t$ with $0 \in X_i+t$
and $w(X_i+t)\le L/4$. 
So we have $g_\Lambda^\prime(X_i)=g_\Lambda(X_i)$,
and the contribution of this term to 
$F_\infty(g^\prime_\Lambda)(X)$ 
will be the same 
as the contribution of this term to $F_\Lambda(g_\Lambda)(X)$.  
So the difference between $F_\infty(g^\prime_\Lambda)(X)$ and
$F_\Lambda(g_\Lambda)(X)$ comes from terms with $\sum_i w(X_i) > L/4$. 
Their total contribution to \reff{fixdif} is of order $\weight^{L/4}$.
Combining this with \reff{latter}, we have shown 
\reff{fixdif} is of order $\weight^{L/4}$. 
If we apply the contraction mapping theorem with the norm without the 
weighting factor $\weight^{w(X)}$, this implies 
\be
\sum_{X \ni 0} |g^\prime_\Lambda(X) - g_\infty(X) |
\ee is of order $\weight^{L/4}$.  
And so \reff{expfast} converges to zero like $\weight^{L/4}$ in the 
infinite volume limit.

\section{Excited states}
\label{excited1}

In this section we study the low-lying excited 
states with one quasiparticle for the model defined by \reff{eqb}
when $\eps$ is small. 
When $\epsilon$ is $0$, these states for the Hamiltonian 
\reff{eq_orig_ham} have all spins pointing in the $z$-direction except 
for one spin pointing in the opposite direction.
Thus the energy of the first excited state is $|\L|$-fold degenerate
($|\L|$ is the number of sites). When $\epsilon$ is small but not zero, 
this degenerate eigenvalue should open up into a band of continuous
spectrum in the infinite volume limit. The goal is to get a convergent 
expansion for the eigenfunctions corresponding to this continuous 
spectrum. This looks impossible since there is no gap around 
these eigenvalues. However, these excited states are uniquely 
determined by their momentum. If $T_l$ denotes translation
by $l$, then an eigenstate with momentum $k$ satisfies
\be
T_l | \Psi > = e^{-ikl} | \Psi > .
\label{momentum}
\ee
(For $d>1$, $k$ and $l$ are vectors, and $kl$ will denote their dot
product.)
If you look in the subspace with momentum $k$, then the first excited state
has an isolated eigenvalue. We will show that by building into the 
wave function the constraint that it has momentum $k$, one 
can get a convergent expansion for the eigenfunction. 
One of the consequences of our expansion is the following 
theorem about the energy dispersion relation. 

We will always work on a hyper-rectangle:
\be
\Lambda=\{(i_1,\cdots,i_d): 1 \le i_j \le L_j, j=1,2,\cdots d\}
\ee
where $L_i$ is the length in the $i$th direction.
The possible momenta are given by 
$k=(k_1,\cdots,k_d)$
where $k_i=\frac{2\pi n_i}{L_i}$  and $n_i=0,1,....,L_i-1$.
We denote the ground state energy for a volume $\L$ by $E_0$.
We let $E_1(k)$ denote the  
lowest eigenvalue in the subspace of momentum $k$ if $k \ne 0$ and 
the next to lowest eigenvalue if $k = 0$. 
(We think of this as the energy of the state with one quasiparticle 
with momentum $k$.)
These eigenvalues depend on the volume $\L$. When we wish to emphasize 
this dependence, we will write them as  
$E_0^\L$ and $E_1^\L(k)$.
The dispersion relation is 
\be
\Delta^\Lambda(k)=E_1^\L(k)-E_0^\L .
\ee
When $\epsilon=0$,  $\Delta^\Lambda(k)=2$ for all $k$. 
For small nonzero $\epsilon$ we expect $\Delta^\Lambda(k)$ to equal
$2$ plus a function that is of order $\epsilon$. 
The following theorem essentially says that this function has an
infinite volume limit, and this limit has a convergent expansion.
We state the theorem in terms of the Fourier coefficients of 
$\Delta^\Lambda(k)$ rather than the function itself since 
$\Delta^\Lambda(k)$ is defined for different values of $k$ 
for different $\Lambda$.

\begin{theorem}
There exists an $\eps_0 >0$ such that for all $|\eps| < \eps_0$
the following is true.
Write $\Delta^\Lambda(k)$ as a Fourier series:
\be 
\Delta^\Lambda(k)= 2 + \sum_{s \in \Lambda} e^\Lambda_s \, e^{iks} .
\label{eq_quasi}
\ee 
There are coefficients $e_s$, defined for all $s \in \zed^d$, such that  
for any sequence of finite volumes which are hyper-rectangles and 
converge to $\zed^d$, 
the coefficients $e^\Lambda_s$ converge uniformly to $e_s$.
The coefficients $e^\Lambda_0$ and $e_0$ are at least first order 
in $\epsilon$, and $e^\Lambda_s$ and $e_s$ are at least of 
order $\epsilon^{|s|}$ in the sense that there is a constant $c$ such that 
\be 
|e_s| \le (c |\eps|)^{|s|} .
\ee
Thus the infinite volume dispersion relation may be defined by 
\be 
\Delta(k)= 2 + \sum_{s \in \zed^d} e_s e^{iks} .
\ee
\label{thm_excited}
\end{theorem}

\bigskip
\no {\bf Remark:} The bounds on the $e_s$ are only upper bounds. 
Some of the $e_s$ may be higher order in $\epsilon$ than the 
theorem indicates. 
\bigskip

We start the proof by considering what states with momentum $k$ look like.  
A function $\psi(\s)$ has momentum $k$ if 
\be 
T_j \psi(\s)=e^{-ikj} \psi(\s) .
\label{basisk}
\ee
If $X$ is any subset of the lattice, then we can form such a 
function by letting 
\be
\phi_{X,k}(\s)=\sum_l e^{ikl} \s(X+l) ,
\ee
where $X+l$ is the set of sites $\{i+l:i \in X\}$. 
The above sum is over sites in $\Lambda$. All subsequent sums in 
this section will be over sites in $\Lambda$ unless otherwise indicated.
We claim that these functions span the subspace of momentum $k$. 
It suffices to show that if we consider all $k$, then these 
functions span the entire state space. Since
\be
\sum_k \phi_{X,k}(\s)= |\Lambda| \s(X) ,
\ee
the span of the functions $\phi_{X,k}(\s)$ includes the 
functions $\s(X)$ for all $X$. The latter form a  
basis for the state space, so the $\phi_{X,k}$ span the 
state space. 

While these functions span the subspace of momentum $k$, they are 
not linearly independent.   
For some choices of $X$ and $k$, 
$\phi_{X,k}(\s)$ will be zero. For example, if $X=\emptyset$ or 
$\Lambda$, then the function is zero unless $k=0$. 
Furthermore, 
\be 
\phi_{X+t,k}(\s) = e^{-ikt} \phi_{X,k}(\s) .
\label{eqnotind}
\ee
So if two $X$'s are translates of each other, then the 
corresponding functions will be the same up to a multiplicative constant. 
Define $X$ and $Y$ to be translationally equivalent if $X=Y+n$ for 
some $n \in \Lambda$. Then the subsets of $\Lambda$ can be partitioned
into equivalence classes. Pick one set from each equivalence class 
and let $\xx$ be the resulting collection of sets. Every subset of $\Lambda$ 
can be written in the form $X + n $  for some $X \in \xx$.
While $X$ is uniquely determined, $n$ is not. (For example, consider
$X=\emptyset$ or $\Lambda$.)
If we consider the collection 
$\phi_{X,k}$ for $X \in \xx$, 
then this collection spans the subspace of momentum $k$. 

Now suppose that for each $k$ we have an eigenstate $\psi_k(\s)$ 
with momentum $k$. Let $\psi(\s)$ denote the groundstate wavefunction,
\reff{eqc}, obtained in Section \ref{ground1}.
Then $\psi_k(\s)/\psi(\s)$ also has momentum $k$. So it can be 
written in the form 
\be 
{ \psi_k(\s) \over \psi(\s)}
= \sum_{X \in \xx} c(X,k) \, \phi_{X,k}(\s)
\ee
for some coefficients $c(X,k)$, which depend on $k$.  
For each $X$, $c(X,k)$ has a Fourier series 
\be 
c(X,k) = \sum_n e^{-ikn} \, e(X,n) .
\ee
(The sum over $n$ is over the sites in $\Lambda$.)
Then 
\be 
\psi_k(\s) = \psi(\s) \, \sum_{X \in \xx}
\sum_n e^{-ikn} \, e(X,n) \, \phi_{X,k}(\s) .
\ee
Using \reff{eqnotind} this becomes 
\be 
\psi_k(\s) = \psi(\s) \, \sum_{X \in \xx}
\sum_n e(X,n) \, \phi_{X+n,k}(\s)
 = \psi(\s) \, \sum_X e(X) \, \phi_{X,k}(\s) ,
\ee
where the coefficients $e(X)$ are given by 
\be 
e(X)= \sum_{Y \in \xx,n: Y+n=X} e(Y,n) .
\ee
We now have 
\be 
\psi_k(\s)= \psi(\s) \, \sum_l e^{ikl} \, \sum_X e(X) \, \s(X+l) .
\label{eqwave}
\ee
If we define 
\be
\phi(\s) = \sum_X e(X) \s(X) 
\ee
then 
\be
\psi_k(\s) = \psi(\s) \sum_l e^{ikl} \, T_l \phi(\s) .
\label{ansatz}
\ee
Note that the only momentum dependence in the above is in the 
explicit factor of $e^{ikl}$. 

When $\epsilon=0$, the lowest excited states with momentum $k$ 
of the Hamiltonian $H$ defined in \reff{eqb} are
\be
\psi_k(\s) = \sum_l e^{ikl} \, \sigma_l .
\ee
In the above formulation this corresponds to 
$c(\{0\},k)=1$ for all $k$ and 
$c(X,k)=0$ for all $k$ and all $X \in \xx$ other than $\{0\}$.  
This is equivalent to $e(\{0\})=1$ and $e(X)=0$ for 
all $X$ other than $\{0\}$.  
For nonzero $\epsilon$ we can normalize the eigenstates so that 
$c(\{0\},k)=1$ for all $k$. This is equivalent to taking 
$e(\{0\})=1$ and $e(\{l\})=0$ for $l \ne 0$. 
We will show that the other $e(X)$ are small and localized 
near $0$ when $\epsilon$ is small. 
As we showed in the last section, the ground state wave function 
$\psi(\s)$ can be written as
$$\psi(\s) = \exp[-\half \sum_X g(X) \s(X)]$$
The ground state is translationally invariant, so $g(X+l)=g(X)$. 

The action of the Hamiltonian on the wavefunction \reff{eqwave} is 
\begin{eqnarray*}
(H \psi_k)(\s) &&= \psi(\s) \sum_l e^{ikl} \\
&& \biggl[ - \sum_j \exp(\sum_{Y \ni j} g(Y) \s(Y))  
  \Bigl[ \sum_{X \not\ni j-l} e(X) \s(X+l) 
  - \sum_{X \ni j-l} e(X) \s(X+l) \Bigr] \\
  && + \epsilon \sum_{<ij>} \, \s_i \s_j
  \sum_X e(X) \s(X+l) \biggr] .\\
\end{eqnarray*}
The above needs to equal 
$$E_1(k) \psi_k(\s) = E_1(k) \psi(\s) \sum_l e^{ikl} 
  \sum_X e(X) \s(X+l) .
$$
Cancelling the common factor of $\psi(\s)$, the Schrodinger
equation for the excited state becomes 
\bea
\sum_l e^{ikl} 
  && \biggl[ - \sum_j \exp(\sum_{Y \ni j} g(Y) \s(Y))  
  [ \sum_{X \not\ni j-l} e(X) \s(X+l) 
  - \sum_{X \ni j-l} e(X) \s(X+l)] \nonumber \\
  && + \epsilon \sum_{<ij>} \, \s_i \s_j
  \sum_X e(X) \s(X+l)  
   - E_1(k) \sum_X e(X) \s(X+l) \biggr] =0 .
\label{excitedse}
\eea
Recall that the ground state wave function satisfies
\be
  - \sum_j \exp(\sum_{Y \ni j} g(Y) \s(Y))  
 + \epsilon \sum_{<ij>} \, \s_i \s_j = E_0 .
\ee
Multiplying this equation by $\sum_l e^{ikl} \sum_X e(X) \s(X+l)$ and 
subtracting the result from eq.  
\reff{excitedse}, we have
\bea
&& 2 \sum_l e^{ikl} \sum_j \exp(\sum_{Y \ni j} g(Y) \s(Y))  
  \sum_{X \ni j-l} e(X) \s(X+l) \nonumber \\
= && \sum_l e^{ikl} [E_1(k)-E_0] \sum_X  e(X) \s(X+l) .
\label{exse}
\eea
Define $h(Y)$ by 
\be 
\exp(\sum_{Y \ni 0} g(Y) \s(Y))  = 1+ \sum_Y h(Y) \s(Y) .
\ee
Explicitly,
\be 
h(Y) = \sum_{n=1}^\infty {1 \over n !} 
\sum_{Y_1,\cdots,Y_n: 0 \in Y_j, 
\atop{Y_1 \d \cdots \d Y_n=Y} } 
g(Y_1) \cdots g(Y_n) .
\label{hdef}
\ee
Note that $h(Y)$ is localized near the origin.
Using the translation invariance of the $g(Y)$, we have
\be
\exp(\sum_{Y \ni j} g(Y) \s(Y))  = 1+ \sum_Y h(Y) \s(Y+j) .
\ee
Inserting this in \reff{exse}
\bea
&& 2 \sum_l e^{ikl} 
  \sum_j  [1+ \sum_Y h(Y) \s(Y+j)]
   \sum_{X \ni j-l} e(X) \s(X+l) \nonumber \\
= && \sum_l e^{ikl} [E_1(k)-E_0] \sum_X e(X) \s(X+l).
\label{exseb}
\eea
The change of variables $j \rightarrow j+l$ changes this to 
\bea
&& 2 \sum_l e^{ikl} \sum_j 
 [1+ \sum_Y h(Y) \s(Y+j+l)]
  \sum_{X \ni j} e(X) \s(X+l) \nonumber \\
= && \sum_l e^{ikl} [E_1(k)-E_0] \sum_X e(X) \s(X+l). 
\eea
Using 
\be
\sum_j \sum_{X \ni j} e(X) \s(X+l)
= \sum_X |X| e(X) \s(X+l)
\ee
this becomes 
\bea
&& 
2 \sum_l e^{ikl}  \sum_X  |X| e(X) \s(X+l)
+ 2 \sum_l e^{ikl} \sum_Y \sum_X \sum_{j \in X} 
h(Y) e(X) \s([(Y+j) \d X]+l) 
\nonumber \\
= && \sum_l e^{ikl} [E_1(k)-E_0] \sum_X e(X) \s(X+l).  
\label{exsec}
\eea

We can write $E_1(k)-E_0$ as a Fourier series in $k$. Since this 
quantity equals $2$ when $\epsilon=0$, it is natural to write it in 
the form   
\be
E_1(k)-E_0 = 2 + \sum_m e^{-ikm} e_m .
\label{eq_fourier}
\ee
Inserting this in \reff{exsec} and making the change of variables 
$l \ra l+m$ and then $X \ra X-m$ in the second term, the right side of 
\reff{exsec} becomes
\be
2 \sum_l e^{ikl} \sum_X e(X) \s(X+l) 
+ \sum_{l,m} e^{ikl} e_m \sum_X e(X-m) \s(X+l) 
\ee
Thus eq. \reff{exsec} is of the form 
\be 
\sum_l e^{ikl} \sum_X f(X) \s(X+l) =0
\label{eqgen}
\ee
where 
\be
f(X) = 2 |X| e(X) 
+ 2 \sum_{Y,Z,j \in Z:(Y+j) \d Z=X}  h(Y) e(Z) 
- 2 e(X) - \sum_m e_m e(X-m) .
\label{eqallk}
\ee
Obviously, if $f(X)=0$ for all $X$, then eq. \reff{eqgen} holds. 
We will argue that the converse is true. 

We rewrite \reff{eqgen} as 
\be 
\sum_l e^{ikl} \sum_X f(X-l) \s(X) =0 .
\label{eqgenb}
\ee
This must hold for all $\s$, so for all $X$ and $k$ 
we must have
\be 
\sum_l e^{ikl} f(X-l) =0 .
\ee
Fix an $X$ and regard $f(X-l)$ as a function of $l$. 
Since the above equation holds for all $k$, 
we must have $f(X-l)=0$ for all $l$. 
This proves the converse, and so 
eq. \reff{eqgen} is equivalent to $f(X)=0$ for all $X$. 

We now assume that $e(\{0\})=1$ and $e(\{l\})=0$ for all $l \ne 0$. 
A priori there is no reason that a solution with these properties must
exist, but we will show that it does. 
As we saw before this is essentially a normalization condition on 
the eigenstates.
If $X=\{s\}$, then the $2 |X| e(X)$ and $2 e(X)$ terms in $f(X)$ 
cancel. For such an $X$, $e(X-m)$ is 
zero except when $m=s$, in which case it 
is $1$. So for such $X$, $f(X)=0$ becomes 
\be
e_s = 2 \sum_{Y,Z,j \in Z: (Y+j) \d Z = \{s\}}  h(Y) e(Z) .
\label{es}
\ee
For $X$ with $|X| \ne 1$ (including the case of $X= \emptyset$), 
the equation $f(X)=0$ can be written as 
\be 
e(X) = { 1 \over 2(|X|-1)} 
\left[ - 2 \sum_{Y,Z,j \in Z: (Y+j) \d Z = X}  h(Y) e(Z) 
+  \sum_m e_m e(X-m) \right] .
\label{ex}
\ee

We will show these equations have a solution by writing them as a 
fixed point equation. 
Consider the set of variables 
\be e := \{e(X): |X| \ne 1\} \cup \{e_s: s \in \L \} .
\ee
Equations \reff{es} and \reff{ex} form a fixed point equation for $e$
\be
\fp(e)=e .
\label{map}
\ee

Let us introduce the norm
\be
||e|| := \sum_s |e_s|\,\weight^{- |s|} + 2 \sum_{X: |X| \ne 1} |e(X)| ||X|-1| 
\weight^{-w_0(X)} .
\label{norm}
\ee
The factor of 2 is included merely for convenience.
Here $w_0(X)$ denotes the number of bonds in the smallest
connected set of bonds containing $X$ and the origin, and 
$|s|$ is the $l^1$ norm of the site $s$. Note that $|s|=w_0(\{s\})$.
If $X = X_1 \d \cdots \d X_n$, then
\be
w_0(X) \le w_0(X_1) + \cdots + w_0(X_n) .
\label{wx}
\ee

We prove that the fixed point equation for $e$ has a solution 
by using the contraction mapping theorem as we did in the previous 
section. We must show that there is a $\delta>0$ such that 
\be
||\fp(e)-\fp(e')|| \le \frac{1}{2} ||e - e'|| \quad {\hbox{for}} \quad
||e||, ||e'|| \le \delta , 
\label{exboundone}
\ee
\be
||\fp(e)|| \le \delta \quad {\hbox{for}} \quad
||e|| \le \delta .
\label{exboundtwo}
\ee

To verify \reff{exboundone}, we use  
\reff{es} and \reff{ex} to see that 
\bea
&& ||\fp(e) - \fp(e')|| 
\nonumber \\
&& \le  \sum_{X:|X| \ne 1} \weight^{-w_0(X)}
  \left[ 2 \sum_{Y,Z,j \in Z: (Y+j) \d Z = X} | h(Y) [e(Z)-e^\prime(Z)] |
+ \sum_s | e_s e(X-s) - e^\prime_s e^\prime(X-s)| \right]
\nonumber \\
 && + 2 \sum_l \weight^{-|l|} \sum_{Y,Z,j \in Z: (Y+j) \d Z = \{l\}}  
  | h(Y) [e(Z)-e^\prime(Z)] | .
\eea
For a site $l$, we have $w_0(\{l\})=|l|$, so the above 
\bea
&& = 2  \sum_{Y,Z,j \in Z} \weight^{-w_0((Y+j) \d Z)}
| h(Y) [e(Z)-e^\prime(Z)]|
\nonumber \\
+ && \sum_s  \sum_{X:|X| \ne 1} \weight^{-w_0(X)} 
| e_s e(X-s) - e^\prime_s e^\prime(X-s)| .
\label{exboundonea}
\eea
We claim that for $j \in Z$ we have $w_0((Y+j) \d Z) \le w_0(Y) + w_0(Z)$. 
To see this, let $A$ and $B$ be connected sets of bonds containing the 
origin such that $Y \subset A$, $Z \subset B$
and $|A|=w_0(Y)$ and $|B|=w_0(Z)$. 
Since the origin is in $A$, the site $j$ is in $A+j$.
The site $j$ is also in $Z$ and so is in $B$.
Since each of $A+j$ and $B$ is connected and they both contain $j$, 
their union is connected. 
So if we define $C=(A+j) \cup B$, 
then $C$ is a connected set of bonds which contains the origin and
$(Y+j) \d Z$. So 
\be 
w_0((Y+j) \d Z) \le |C| \le |A+j| + |B| = |A| + |B| = w_0(Y) + w_0(Z) .
\ee
We also have $w_0(X) \le w_0(X-s) +|s|$. The sum over $j$ in \reff{exboundonea}
then produces a factor of $|Z|$. So \reff{exboundonea} is 
\bea
&& \le 2  \sum_{Y,Z} \weight^{-w_0(Y)-w_0(Z)} |Z| | h(Y) 
[e(Z) - e^\prime(Z)] |
\nonumber \\
&& + \sum_s  \sum_{X:|X| \ne 1} \weight^{-w_0(X-s)-|s|}
| e_s e(X-s) - e^\prime_s e^\prime(X-s)| .
\label{eqcbounda}
\eea
Since $e(Z)=e^\prime(Z)$ if $|Z|=1$, and $|Z| \le 2 ||Z|-1|$
if $|Z| \ne 1$, we have 
\be
\sum_Z \weight^{-w_0(Z)} |Z| | e(Z) - e^\prime(Z)| 
\le \sum_{Z:|Z| \ne 1} \weight^{-w_0(Z)} 2 ||Z|-1| | e(Z) - e^\prime(Z)| 
\le ||e-e^\prime|| .
\ee
So the first of the two terms in \reff{eqcbounda} is bounded by
\be
2 \sum_Y \weight^{-w_0(Y)} \, |h(Y)| \, ||e-e^\prime|| . \label{eqboundz}
\ee
After the change of variables $X \ra X+s$, the second term in \reff{eqcbounda} 
equals 
\bea 
&& \sum_s  \sum_{X:|X| \ne 1} \weight^{-w_0(X)-|s|}
| e_s e(X) - e^\prime_s e^\prime(X)| 
\nonumber \\
&& \le 
 \sum_s  \sum_{X:|X| \ne 1} \weight^{-w_0(X)-|s|}
[ | e_s e(X) - e_s e^\prime(X)| 
+ | e_s e^\prime(X) - e^\prime_s e^\prime(X)| ]
\nonumber \\
&& \le 
{ 1 \over 2} ||e|| \, ||e-e^\prime||
+ { 1 \over 2} ||e-e^\prime|| \, ||e^\prime||
\le \delta \, ||e-e^\prime|| , 
\label{eqcboundb}
\eea
where we have used the triangle inequality 
and the fact that $||e||$ and $||e'||$ are no greater than $\delta$.
Combining \reff{eqboundz} and \reff{eqcboundb}, we have 
\be
||\fp(e) - \fp(e')||\le K \, ||e - e'|| ,
\label{f1e}
\ee
where
\be
K = 2 \sum_Y \weight^{-w_0(Y)} |h(Y)| + \delta .
\label{k1}
\ee
{From} \reff{hdef} and \reff{wx}
\be 
\sum_Y \weight^{-w_0(Y)} |h(Y)| \le \sum_{n=1}^\infty {1 \over n !} 
\sum_{Y_1 \ni 0,\cdots,Y_n \ni 0}
\weight^{-w_0(Y_1) - \cdots - w_0(Y_n)}
|g(Y_1) \cdots g(Y_n)|
= e^{||g||}-1 .
\ee
So 
\be
K \le 2(e^{||g||} -1) + \delta .
\ee
If $\delta$ and $\epsilon$ are small enough, then $K \le 1/2$. 

To prove \reff{exboundtwo}, we use \reff{es} and \reff{ex} to compute
$\fp(0)$. Note that $e=0$ means that $e(X)=0$ for all $X$ except
$X=\{0\}$. We always have $e(\{0\})=1$.
Letting $e^\prime$ denote $F(0)$, we have 
\be
e^\prime_l = 2 \sum_{Y: Y \d \{0\} = \{l\}}  h(Y) 
\ee
and for $X$ with $|X| \ne 1$ (including the case of $X= \emptyset$), 
\be 
e^\prime(X) = - { 1 \over (|X|-1)} 
 \sum_{Y: Y \d \{0\} = X}  h(Y) .
\ee
Thus
\be
||\fp(0)|| \le 2 \sum_Y \weight^{-w_0(Y)} |h(Y)| \le 2(e^{||g||}-1) .
\ee
Hence, for $||e|| \le \delta$
and $\epsilon$ small enough that 
$ 2(e^{||g||}-1) < \delta/2 $, 
\be
||\fp(e)|| \le ||\fp(e)-\fp(0)|| + ||\fp(0)||
\le {1 \over 2} ||e|| + ||\fp(0)|| \le \delta .
\ee
This completes the proof that the fixed point equation has a solution.

We still must prove 
that for each $k$, the eigenstate we have constructed has the 
lowest eigenvalue in the subspace of momentum $k$ if $k \ne 0$ and 
has the next to lowest eigenvalue if $k = 0$. When $\epsilon=0$ we
know that this is true. So if we can prove that within the subspace 
of momentum $k$, our eigenstate is never degenerate for 
$|\eps| \le \eps_0$, then we are done. We will use the local 
uniqueness of solutions to the fixed point equation to prove this.

Let $\psi_k(\s)$ continue to denote 
the eigenfunctions constructed above with eigenvalues $E_1(k)$. 
Suppose that for some $k_0$ there is another eigenfunction
$\psi^\prime(\s)$ with momentum $k_0$ and eigenvalue $E_1(k_0)$.
We can write it in the form 
\be 
\psi^\prime(\s) = \psi(\s) \, \sum_{X \in \xx} c^\prime(X) \, 
\phi_{X,k_0}(\s) .
\ee
For small $\alpha$ we define 
\be 
\psi^\alpha_k(\s) = \psi_k(\s)
\ee
for $k \ne k_0$ and 
\be 
\psi^\alpha_{k_0}(\s) = {\psi_{k_0}(\s) + \alpha \psi^\prime(\s)
\over 1 + \alpha c^\prime(\{0\})} .
\ee
(We need only consider small $\alpha$, so we are not dividing by zero in 
the above.)
Then for all small $\alpha$, $\psi^\alpha_k(\s)$ is an eigenstate of $H$ 
with eigenvalue $E_1(k)$ and momentum $k$ for all $k$. 
These states can be written in the form
\be 
\psi_k^\alpha(\s) = \psi(\s) \, \sum_{X \in \xx} c_\alpha(X,k) 
\, \phi_{X,k}(\s) ,
\ee
where $c_\alpha(X,k)$ is just $c(X,k)$ for $k \ne k_0$ and 
\be 
c_\alpha(X,k_0) = {c(X,k_0) + \alpha c^\prime(X,k_0)
\over 1 + \alpha c^\prime(\{0\})} .
\ee
As before we can write this in the form 
\be 
\psi_k^\alpha(\s)  = \psi(\s) \, \sum_X e_\alpha(X) \, \phi_{X,k}(\s) ,
\ee
where
\be 
c_\alpha(X,k) = \sum_n e^{-ikn} \, e_\alpha(X,n) 
\ee
and 
\be 
e_\alpha(X)= \sum_{Y \in \xx,n: Y+n=X} e_\alpha(Y,n) .
\ee
We have normalized things so that $e_\alpha(\{0\})=1$
and $e_\alpha(\{l\})=0$ for $l \ne 0$. 
It follows from what we have done before that for small $\alpha$, 
$e_\alpha(X)$ is a 
solution of the fixed point equation \reff{eqallk}. As $\alpha \ra 0$,
$e_\alpha(X) \ra e(X)$. However, our solution of the fixed point equation
is locally unique by the contraction mapping theorem.
Thus $e_\alpha(X)=e(X)$ for small $\alpha$. This implies that 
the second eigenfunction $\psi^\prime$ is in fact just a multiple 
of $\psi_{k_0}$. This completes the proof that the eigenfunctions 
we have constructed are indeed the lowest excited states.

The existence of the infinite volume limit and the convergence of 
$e^\Lambda_s$ to $e_s$ follow from the estimates of this section and 
the arguments of the previous section. 
In the fixed point equation, \reff{es} and \reff{ex}, $h(Y)$ depends on the 
volume. However, the difference between it and its infinite volume 
limit is exponentially small in $L$. The extension of $e(X)$ to a 
function $e^\prime(X)$ that is defined for all $X$ is simpler 
than the definition of $g^\prime(X)$ in the previous section.
We can simply define $e^\prime(X)$ to be $e(X)$ if $X$ is in $\Lambda$
and $0$ otherwise.
The bounds on the $e_s$ follow from 
our choice of norm \reff{norm} which implies that 
\be
    |e_s| \le \weight^{|s|} ||e||
\ee
and so $e_s$ is of order $|\epsilon|^{|s|}$.   
This completes the proof of Theorem \ref{thm_excited}.
\qed

\medskip

As an application of the theorem, we show that the minimum of 
the dispersion relation occurs at $k=(\pi, \cdots, \pi)$ if 
$\epsilon>0$ and at $k=(0, \cdots, 0)$ if $\epsilon<0$. 
We can use our expansion for the dispersion relation to compute 
the order $\epsilon$ contribution to the dispersion relation.
Consider eq. \reff{es} for the coefficients $e_s$. 
Recall that $e(\{0\})=1$. Otherwise, $e(Z)$ is always at least order
$\epsilon$. For all $Y$, $h(Y)$ is at least order $\epsilon$. 
So $e_s$ can get an order $\epsilon$ contribution from the sum in 
\reff{es} only for $Z=\{0\}$. In this case $j$ must be zero. The constraint
on the sum is $Y \d \{0\}=\{s\}$, which is equivalent to $Y = \{0\} \d \{s\}$.
{From} \reff{fixedpt}
we see that the only $Y$'s for which $g(Y)$ has a nonzero contribution
at first order in $\epsilon$ are the sets containing two sites which 
are nearest neighbors. For such sets $g(Y)=\epsilon/2 + O(\epsilon^2)$.  
Thus from \reff{hdef} 
we see that $h(Y)$ has a nonzero contribution at first
order in $\epsilon$ only if $Y$ is a set consisting of two nearest 
neighbors, one of which is the origin. For such sets 
$h(Y)=\epsilon/2 + O(\epsilon^2)$.  
Since $Y = \{0\} \d \{s\}$,  
$e_s$ has a first order contribution only if $s$ is a nearest neighbor of $0$,
i.e., $|s|=1$. In this case eq. \reff{es} gives $e_s=\epsilon + O(\epsilon^2)$.
Thus 
\be 
\Delta(k)= 2 + \epsilon \sum_{s: |s|=1} e^{iks} +
\sum_{s \in \zed^d} e_s^\prime e^{iks} 
=2 + 2 \epsilon \sum_{i=1}^d \cos(k_i) 
+ \sum_{s \in \zed^d} e_s^\prime e^{iks} .
\ee
where the $e_s^\prime$ are all of order $\epsilon^2$, and $k_i$ are the 
components of the vector $k$. 

We consider the case of $\epsilon>0$. 
The argument in the other case is analogous.
Obviously, $2 + 2 \epsilon \sum_{i=1}^d \cos(k_i)$ has its minimum at 
$k=(\pi, \cdots, \pi)$.
To show this is true for $\Delta(k)$
we consider two cases. 
It is easy to see that $k=(\pi, \cdots, \pi)$ is a critical point of 
$\Delta(k)$ and by computing the Hessian that this critical point must be a 
local minimum. So there is a $\delta>0$ so that 
$\Delta(k)$ is greater than its value at $(\pi, \cdots, \pi)$
provided $|k_i-\pi|<\delta$ for all $i$. If any $k_i$ has 
$|k_i-\pi| \ge \delta$, then
$2 + 2 \epsilon \sum_{i=1}^d \cos(k_i)$ is greater than its minimum 
value by at least $O(\delta^2)$. The difference between 
$2 + 2 \epsilon \sum_{i=1}^d \cos(k_i)$ and $\Delta(k)$ is $O(\epsilon^2)$,
so $|k_i-\pi| \ge \delta$ implies
$\Delta(k)$ is greater than its value at  $k=(\pi, \cdots, \pi)$ provided 
$\epsilon$ is small compared to $\delta$. 

\medskip
\bigskip
\medskip

\noindent {\bf Acknowledgements:}
ND thanks C. Gruber, N. Macris and M. Salmhofer for 
interesting discussions. TK thanks the D\'epartment de Physique 
and the Institut de Physique Th\'eorique of the Ecole Polytechnique 
F\'ed\'erale de Lausanne for their hospitality during a visit when 
much of this work was done. He also acknowledges the support of the 
National Science Foundation (DMS-9970608).

\end{document}